# pH-dependent Response of Coiled Coils: A Coarse-Grained Molecular Simulation Study [*]

Marta Enciso, Christof Schütte, and Luigi Delle Site

*Institute for Mathematics, Freie Universität Berlin, Germany*

In a recent work we proposed a coarse-grained methodology for studying the response of peptides when simulated at different values of pH; in this work we extend the methodology to analyze the pH-dependent behavior of coiled coils. This protein structure presents a remarkable chain stiffness and is formed by two or more long helical peptides that are interacting like the strands of a rope. Chain length and rigidity are the key aspects needed to extend previous peptide interaction potentials to this particular case; however the original model is naturally recovered when the length or the ridigity of the simulated chain are reduced. We apply the model and discuss results for two cases: (a) the folding/unfolding transition of a generic coiled coil as a function of pH; (b) behavior of a specific sequence as a function of the acidity conditions. In this latter case results are compared with experimental data from the literature in order to comment about the consistency of our approach.

---

# I. INTRODUCTION

The description of the relation between external stimuli and their corresponding system response is one of the mandatory questions to understand how nature works. In the case of proteins and peptides, environmental conditions (such as temperature, concentration, pH, ionic strength) strongly modify the folding and aggregation behavior [1]. This aspect is of major importance for understanding processes like folding diseases (like Huntington or Alzheimer) or can be used to design peptide-based medications [2, 3]. However, it also arouses a more fundamental interest about the physical principles involved; these can be studied and eventually understood through theory and simulation [4].

Among many different factors that can affect proteins, pH plays a preponderant role. A large number of amino acids (proteins' building blocks) are highly sensitive to pH, e.g. they can change their protonation state according to the media acidity. Their electrostatic interactions are modified accordingly and, because of that, the overall behavior of peptides and proteins may be affected [5]. The presence of multiple binding sites constitutes a theoretical and computational challenge and its proper description became a field of high interest for the scientific community [6, 7].

Given the complexity of these biological systems and the corresponding computational price, coarse-graining approaches seems a valuable strategy to unveil the role of pH for an extended and relevant class of systems. Physically consistent coarse-grained strategies can be developed according to different modeling schemes; our modeling idea follows the approach based on the principle of "consistency across the scales" [8, 9] by which the reduction of the number of degrees of freedom is such that minimizing structural and thermodynamic artifacts (having higher resolution data as a reference) is the key priority. For the problem of constant-pH methods so far only full-atom approaches have been employed [6, 10, 11], however a coarse-grained approach to the problem has been proposed only recently by us [12].

In our previous work we presented a general methodology to simulate constant-pH conditions at a coarse-grained level. In addition, we introduced an interaction potential for peptides, based on an extensive study of peptides of mainstream features (chain length, global charge, amino acid composition, etc.), which were taken from a broad protein database [13]. We carried out thorough pH-dependent tests on our full model for some relevant cases, showing a clear consistency with experiments and full-atom simulations [12] (i.e. consistency across the scales).

The general principles of our pH-dependent modeling strategy can be applied (in general) to any situation with a coarse-grained resolution, while the transferability of a specific model is limited (in general) to a specific class of systems. For example, peptide coarse-grained models are usually designed to work at common conditions (around physiological temperature conditions, using peptides of average length, charge, etc.), however there are some cases where it may be

4interesting to study pH-dependent effects in not-so-usual situations; this is the case of coiled-coils.

Coiled-coils are one of the simplest super-secondary protein structures, consisting of two or more $\alpha$-helices wrapped around each other in a superhelical twist, as we show in Figure 1(a). Coiled-coils are present in many protein domains and can form fibers, ion channels and pores. They present multiple possibilities for functionalisation and applications [14–16] and play a relevant role in biological processes like HIV infection, DNA transcription, etc. [15, 17]. Most importantly from the modeling point of view, their structural simplicity and regularity makes them an ideal target for model testing [18, 19].

Unlike other peptides, coiled-coils are usually formed by particularly long peptide chains (from 15 up to many tens or even hundreds of amino acids), much longer than the usual peptide length (10-15 amino acids). However, they do not have the characteristic complex structure of a full protein of comparable size, actually something more similar to a peptide; thus they sort of "bridge" between proteins and peptides because their structure is still very simple but the system size and complexity are significantly larger.

This structural simplicity is due to the unusual stiffness of such long chains (frequently described as rod-like fibers) [20]. For this reason, this is an ideal test system to refine our previous peptide interaction model and extend it to the specific features of this protein family for the study of the response to different pH values.

The effect of pH in coiled-coils is crucial for its biological role, such as apoptosis in cancer cells [21], and its industrial applications (as biosensors, hydrogels, agents for drug delivery, etc.) [22], where they are exposed to changeable pH environments. Different pH-dependent coiled coils have been studied and even designed recently, with different responses towards the media acidity. Depending on the particularities of the peptide sequence, it has been experimentally found that a decrease in the pH conditions may increase the stability, decrease it or keep it the same [16, 23–26]. Our aim is to apply our coarse-grained strategy to reproduce in molecular simulations reasonable response to different values of pH; this would broaden the applicability of our model to a larger class of systems with similar characteristics, i.e. long and stiff helical chains.

This paper is organized as follows. First, we present a methodological section with technical details such as a description of the pH-dependent coarse-grained model, the length-dependent study and the simulation setup. In section IV we show our simulation results about the influence of the chain length in coiled-coil structures, followed by some concluding remarks.

## II. TECHNICAL DETAILS

In this section we describe the main technical details we have used to generate the results presented here. Those concerning the previous coarse-grained pH-dependent model are explained in dept somewhere else [12].

We describe the peptide system by an off-lattice bead and stick representation with a single center of interaction per amino acid (placed at the $\alpha$-carbon position), with implicit solvent. Consecutive beads in the same chain are placed at a constant distance of 0.38 nm (corresponding to a *trans* peptide bond). Fig.2 pictorially shows the basic chain model, while the explicit form of the interaction potentials is given later on in this section.

We have then carried out simulations by means of a Parallel Tempering Monte Carlo in-house software, parallelized with OpenMP for higher performance. We have performed single-chain and multi-chain numerical experiments using periodic boundary conditions, where each of our simulations presents 24-40 temperatures. Simulations start from a completely extended conformation for each chain and consists of $8 \cdot 10^6$ Monte Carlo cycles at every temperature after $3 \cdot 10^6$ equilibration cycles (enough to guarantee convergence). In each cycle, every bead of the system is subjected to a trial Monte Carlo move. In order to sample the conformational space as efficiently as possible, we have implemented three individual-bead movements (spike, rotation or translation); besides, each chain is allowed to rotate or translate. For each system, 3-5 independent runs have been carried out.

pH has been modeled using discrete protonation states that change during the simulation according to a Monte Carlo scheme [12]. In each of these "pH moves", a protonable site is randomly selected and its protonation state changed. Then, the energetic difference between the current and previous configuration is calculated and the "movement" is accepted or rejected according to the detailed balance condition, that depends on the response of the particular amino acid towards protonation (ie. its equilibrium acidity constant) and on the effect of pH on the overall interactions. This effect is explained at the end of the next section which gives an overview of the driving interactions in our system.

### A. The pH-dependent coarse-grained potential for short chains

In our basic coarse-grained model the main peptide driving forces are considered independently, and then summed in the global potential, following this expression:

$$E = \omega^{hb} \ E^{hb} + \omega^{hp} \ E^{hp} + \omega^{stiff} \ E^{stiff} + \omega^{elec} \ E^{elec}$$

The main contributions are hydrogen bonds, $E^{hb}$, electrostatic interactions, $E^{elec}$ and hydrophobic interactions, $E^{hp}$; we have also included a "stiffness term". The details of these terms are explained somewhere else [12, 27, 28], so we only describe here their main features.

The hydrogen bond interaction, $E^{hb}$, is applied between any pair of residues $i$ and $j$ (where $j > i + 2$ and $j \geq i + 4$). Its energy calculation consists of two steps. First, we check three geometrical restrictions (namely, the length of the tentative hydrogen bond between beads; the orientation between the auxiliary vectors; and the relative orientation between those auxiliary vectors and the tentative hydrogen bond), which have been schematically plotted in Figure 2. If the values of the former restrictions fall within certain limits, a step-wise potential is applied. Acceptable ranges and potential strength differ depending on the kind of hydrogen bond (either local/helical or non-local/$\beta$-type). For more details see Ref. [27].

The hydrophobic interaction, $E^{hp}$, is modeled by a 10-12 Lennard-Jones potential between pairs of residues. It distinguishes between hydrophobic and polar centers of interaction [28]. Amino acids have been classified into these two categories according to Ref. [29]. We have also introduced a sequence-independent stiffness term, $E^{stiff}$, that controls the chain flexibility and equally favors helical and extended conformations [28, 29]. The stiffness term is computed for each residue in the chain, $i$, and depends on the bond angle among three consecutive beads, $\phi$:

$$E_i^{stiff} = 0.5\ (1 + \cos 3\phi_i) - 1$$

This potential presents minima at $\phi_i = \pm\pi/3$ (matching the most probable angle of $\alpha$-helices) and $\phi_i = \pm\pi$ (most probable angle in extended configurations such as $\beta$-sheets).

So far the effect of the pH as not been counted; in fact, we proved in a previous work that the interactions above are not affected by different values of pH; instead, pH strongly affects electrostatic interactions [12]. In order to model the effect of pH we have taken full atomistic simulations and experimental data and used as a target to model "effective" electrostatic interactions. We have found that our $E^{elec}$, can be successfully modeled by a potential with Yukawa functional form [30]; testing details can be found in Ref. [12].

The balance among the contributions is defined in terms of the following weighting factors: $\omega^{hb} = 9.5$, $\omega^{hp} = 6.5$, $\omega^{stiff} = 7.0$ and $\omega^{elec} = 12.0$. We use energetic units referred to a certain reference state of temperature $T_{ref}$ and energy $E_{ref} = k_B T_{ref}$; if needed, they can be linked to a real experimental state. We have then performed our simulations in terms of these reduced and adimensional units, defined as $T^* = T_{real}/T_{ref}$ and $E^* = E_{real}/k_B T_{ref}$.

### III. COILED-COIL SYSTEM

A coiled coil is a structural motif in proteins, in which several $\alpha$-helices (usually two or three) are wrapped against the others like the strands of a rope. This arrangement is obtained thanks to the presence of a repeated pattern of seven amino acids [17], usually called heptad repeat and labeled *abcdefg*, as we have sketched in Figure 1(b).

It has been found that hydrophobic, polar and charged residues are usually arranged following certain rules [17, 31]. According to these rules, there is a hydrophobic core in positions *a* and *d* (usually occupied by leucines and valines) and charged residues (like lysine and aspartic acid) are frequently placed in positions *e* and *g*, forming interchain electrostatic interactions. Polar residues are often found in the other positions of the heptad, as they are exposed to the solvent.

Following these principles, we have designed a simple algorithm to build coiled coil sequences of any length, and used that sequences accordingly. For instance, the following sequence has been used for a peptide of 17 amino acids: TQIEREVDKIRQEIHKI.

### A. Length-dependent effects in coiled coils

The unusual rigidity of coiled coils is such that they stay helical even in long chains; from the methodological point of view we need to analyze now how our basic potential behaves with these them. For this purpose we have started by simulating just one chain (with a typical coiled coil sequence) in the simulation box, at neutral pH (pH=7). In these single-chain REMC simuations one would expect the formation of a helix at temperatures lower than the transition one (i.e. the stable state of the system should be a helix). We have performed this folding/unfolding test in chains of different lengths, finding the transition temperatures ($T_m^*$) displayed in Table I.

The helicity of each of these systems in the folded region (defined in this case by a temperature $T^* = 0.8\ T_m^*$) is measured by the mean residue ellipticity at 222 nm, which can be computed with the method described in Refs. [32, 33]. We have referred it to its maximum value, $[\theta_{max}]$ (ie. the value that a perfect helix would have). This "relative ellipticity", $[\theta_{rel}]$, is shown in Figure 3(a).

We observe in this Figure that short chains present $[\theta_{rel}] \simeq 0.7$, that is, a 70% helicity. Longer chains exhibit a drastic decrease in the chain ellipticity. This implies that our interaction potential, appropriate for short chains and other types of peptides [12], predicts a drop in the helicity of longer chains. This is reasonable for other sequences, as peptides of these lengths seldom form a unique helix, but more compact structures such as $\beta$-sheets or helical bundles (shorter helical fragments packed within the same chain), fruits of the hydrophobic interactions. The very particular properties of coiled-coils make this finding unsuitable for our current purposes.

In Figure 3(a) we must also notice that the drastic change of behavior occurs at a critical helix length, $l_{crit} = 18$ amino acids. This value define the separation between two different scenarios: short-chain regime, where helical conformations are favored, and the long-chain regime where a high helicity is no longer preferred.

The appearance of distinct properties above a certain peptide length has been found to be quite common in protein and polymer systems [34–37]. It is related to the fact that small peptides have a smoother free energy surface than longer ones [38]. In our case, however, our target is not the change in the system properties, but the refinement of the interaction potential so that coiled coils keep their natural helicity.

Next, we have analyzed some structural properties of our data, again at the same temperature

$T^* = 0.8\ T_m^*$; we have calculated the persistence length with respect to the first bond, for the different chain lengths, see Figure 3(b). This quantity, very common to evaluate the flexibility of peptides or polymers, measures the rigidity of a certain chain (i.e. larger $l_p$ implies a more rigid chain) and can be calculated using the following formula:

$$l_p = \frac{1}{b} \sum_{k=0}^{N-2} <\mathbf{v}_1 \cdot \mathbf{v}_{1+k}>$$

; $b$ is the bond length (fixed at 0.38 nm in our simulations), $N$ is the number of amino acids in the chain and $\mathbf{v}$ is the bond vector between two consecutive residues. We have normalized the persistence length dividing it by the number of residues in the chain (to remove any length dependency) and the bond length.

We can see in Figure 3(b) that the persistence length of our chains clearly drops above the critical length, indicating the presence of compact structures. This would be again an expected result for other kinds of peptides of similar length, but not for long helical chains such as the ones that form coiled coils. In fact, we have checked this point by simulating a 23-residue peptide of PDB code 2DJ9 whose native structure is a $\beta$-hairpin. In this case our model mainly forms compact conformations of low persistence length, compatible with the native state and its sequence. This implies that the original model is valid not only for short chains, but also in those cases where long helical chains are not expected.

If we analyze how the interactions among residues are distributed in short and long chains, we can find out a quantitative picture of the two different scenarios. As an example, we show in Figure 4 these two-dimensional plots for a short peptide (15 residues) and a long one (29 residues). The axis are counters of the amino acids of the chain, and the colored dots indicate interactions, in logarithmic scale, between the residues in the axes. The computation is done as an average over all the sampled configurations at the temperature of interest, and only the stabilizing interactions are plotted in order to make the map clearer.

We see in Figure 4(a) (short chain) that most of the interactions are placed in the surroundings of the square diagonal, ie. correspond to residues close in sequence (local interactions, which stabilize helices). In Figure 4(b), however, the long chain shows a completely different distribution of the interactions. We observe that nearly the whole map presents some color, which means that most of all possible pair-wise interactions have been formed at least momentarily throughout our simulation. This fact matches the drop in the system rigidity at long lengths. Besides, we observe the presence of green and blue colors far away from the plot diagonal (ie. long range interactions), typical of $\beta$-type and compact structures like the ones mentioned in the case of the $\beta$-type peptide.

We have shown, then, that the original potential presents two different regimes depending on the chain length. While short chains behave like $\alpha$-helices, long ones experience a sort of collapse of the structure, characterized by a decrease in the chain stiffness and the presence of multiple long range interactions. In essence this is the key factor to understand why the data used to model short chain are insufficient for deriving a model extendable to coiled coils. In the next section we extend the modeling of short chains to this case based on the results discussed in the current section.

## B. Model improvement: the length-dependent chain stiffness

We have observed in that our model stabilizes quite compact structures when applied to peptides beyond a certain critical length. This view is not desirable in the current case, as our target is the coiled coil, i.e. long $\alpha$-helices twisted around one another. It has been found that this structure is stable because of a remarkable stiffness of each of these chains (often modeled as rod-like fibers in the literature) [20]. Inspired by this finding, we have tuned our interaction potential so that it describes a typical coiled coil behavior.

The potential described in subsection II A already presents a stiffness interaction, so we have modified it according to our purposes. We have included a chain-dependent term that reduces the chain flexibility and favors the helical structure that has been found in experiments, which is characterized by a local angle $\phi$ centered around $\pi/3$. This term has been adapted from the original works of Head-Gordon *et al.* for the case of helical residues [29] and follows this expression:

$$E_i^{stiff,long} = 1.2[(1 - \cos\phi_i) + (1 + \cos 3\phi_i) + \left(1 + \cos\left[\phi_i + \frac{\pi}{4}\right]\right) - 4.8$$

It presents a sinusoidal shape with a global minimum at $\phi_i = \pi/3$. Then, the final stiffness interaction is calculated as a linear combination of the original stiffness term, $E^{stiff,short}$, and the new one:

$$E^{stiff} = \omega_{cc} E_i^{stiff,long} + (1 - \omega_{cc}) E_i^{stiff,long}$$

The weighting factor $\omega_{cc}$ depends on the chain length according to the following expression:

$$\omega_{cc} = \min[\max[0, \frac{l - l_{crit}}{l_{crit}}], 1]$$

$l$ is number of residues of the chain and $l_{crit}$ is the critical length. The very important aspect is that we can naturally recover the original potential in the limit of medium-short chains; note also that the original definition is also valid in the case of long peptides that do not present the high rigidity of a single helix.

Using this improved potential, we have repeated the simulations described in the previous sections. We show in Figure 5 the new ellipticity values of our single-chain simulations, again in folded conditions ($T^* = 0.8\ T_m^*$). We can see that the peptides present a high helicity content regardless their chain length, that is we have (from the structural point of view) successfully refined our model to study this particular case.

The final step in our modeling is the simulation of the full coiled coil structure. We have then simulated two different chains into the same simulation box with our REMC procedure. Depending on the particular concentration conditions, we have found either coiled coils (at medium and high concentrations) or isolated helices (low concentrations), as we expected based on previous works [28]. Then, we have selected proper and relatively high concentration conditions (around 0.15 mmol/L), which have been used in the following section.

It is also importat to note that the formation of coiled coils using our model (ie. the wrapping of the $\alpha$-helices) is a sequence-dependent effect, ie. this model is valid for any long helical chain, while a full coiled coil structure is only observed when the simulated peptides present the proper sequence. To check this point we have carried out simulations at neutral pH for two diferent cases: (1) a generic coiled coil sequence of 29 amino acids (TQIEREVDKIRQEIHKIEQRIQDIKERMQ) and (2) a generic sequence for a helix that does not comply with the coiled coil heptad rules (TEFVQDDIQRYWHRWNEVWQEANQMMRTIR). In the former case, which will be described in detail in the next section, we have observed the predominant formation of helical coiled-coil structures. In the latter we have obtained a mixture of independent $\alpha$-helices and alternative $\beta$-type structures, typical of aggregation and reasonable in high concentration environments such as the simulated one. In fact, the coiled coil is an example of natural self-assembly, ie. how a multi-chain structure keeps its stability without aggregation and collapse [39, 40].

## IV. RESULTS AND DISCUSSION

Coiled coils are peptide structures that combine a relatively simple structure (two interacting $\alpha$-helices) with a relatively large size and, as very often underlined, an important potential in biological and industrial processes, often mediated by environmental factors such as pH. We have recently proposed a sophisticated tool to simulate peptides in different pH media using a coarse-grained resolution and in the previous sections we have refined our interaction potential to properly obtain the distinct characteristics of coiled coil chains; we aim now to analyze the role of pH in these systems.

At this stage we are interested in proving that general trends found in experiments are consistent with the results of our simulation and thus conclude that the basic characteristic of our model are physically well founded; later on, on the basis of this model, one may add specific features to analyze particular cases. In fact, the impact of pH in coiled-coils depends very much on the particularities of each peptide. In some cases it has been experimentally found that an increase in the media acidity leads to a partial unfolding of the coiled coil [23–25]. In other cases low pH conditions exhibit the opposite effect, increasing the stability of the peptides [26]. In some cases it is even possible to control the response towards pH by introducing some mutations into a sequence [16].

Given the sequence dependence in the pH response, we have decided first to make a numerical experiment to understand some generic features which may in the future be even tested by experiments and then apply our model to a specific sequence to check whether our model is consistent with the experimental findings.

For the reason explained above the focus of this section is the study of a generic peptide, designed by a random algorithm that follows the general rules of coiled-coil formation (see subsection III for details). After that we also show results about a fragment of the B-ZIP domain of VBP, which has been experimentally tested in different pH conditions [25].

Regarding the generic peptide, we have taken a long peptide (29 residues, sequence TQIEREVD-KIRQEIHKIEQRIQDIKERMQ) and analyzed the impact of different acidity conditions, from pH=1 (very acid) to pH=13 (very basic). We have performed REMC simulations using two chains in a simulation box; we have also simulated different temperature conditions, in order to track the thermal folding/unfolding equilibrium of the systems.

We have analyzed some properties of our system in the folded region, following the steps explained in section II. They are summarized in Table II. We show in the first column of the Table the transition temperatures in reduced units. Although they all are quite similar, the maximum value is obtained at physiological conditions (pH=7). This hint of higher stability at neutral pH is confirmed by a larger presence of coiled coils at that conditions (second column of Table II). At other pH values coiled coils coexist with alternative structures that have also been characterized for this study. They also have a high helicity content (third column of Table II) but only scarce interchain interactions, ie. they are independent $\alpha$-helices that are no longer forming the full coiled coil structure.

In this case the loss of the multichain structure is a direct consequence of the pH conditions, as the electrostatic balance is severely modified. In the last column of Table II we show for instance the electrostatic energy associated to each pH condition: it is maximum at physiological pH and drops at other pH values, particularly at high pH.

In fact, pH has a direct effect on long range interactions, as it changes the peptide electrostatic properties (charges). These interactions are, together with hydrophobics, the main agent that glue the two peptide chains of a coiled coil (see Figure 1(b)). As it has been reported in other cases [16, 23–26], this effect may be observed at acid conditions or basic ones (such as in this case), depending on the particularities of the sequence.

We have also taken a real sequence for our tests. It is a 29-residue fragment of the B-ZIP domain of VBP which has been studied by Krylov and co-workers in the context of sequence mutations and different environmental conditions [25]. In this particular case it has been found that low pH values lead to a decrease in the protein stability, mainly due to the loss of the electrostatic interactions between glutamic acid and arginine across the interchain interaction surface.

In our case we have carried out folding/unfolding equilibrium simulations at pH=3 and pH=7. We have calculated the system energy at the transition temperature, finding that it is 25% less stable at pH=3 than at physiological pH. This finding correlates well with the experimental enthalpy values, which are 15% lower at pH=2.9 than at pH=7.4 [25]. The source of this energetic difference has been analyzed through two-dimensional energetic maps such as the ones in section II and are shown in Figure 6. Note that the axes include the two chains; the separation between them has been marked with a dashed line.

In the three maps we observe intense interactions along the secondary diagonal, ie. the presence of quite well formed $\alpha$-helices. The differences among them lay in the long range interactions and, more particularly, in the interactions between the two chains (those along the main diagonal). At pH=3 we only observe scarce and weak interactions; at pH=7 the proportion of green/blue colors substantially increase, highlighting the presence of interchain interactions among the involved residues. This is in full consistency with the observations in Ref. [25], where the interaction between amino acids are measured at different pH values through NMR experiments.

Anyway, we can conclude that, according to our model, the effect of pH in coiled coils is direct (ie. it affect the electrostatic interactions themselves) and these electrostatic repulsions drive the destabilization of the full coiled coil structure, this in turn is fully consistent with the current knowledge about coiled coil behavior and thus makes our tests successful from a methodological point of view.

## V. SUMMARY AND CONCLUSIONS

In this paper we extend a previously developed model of peptides that describes the pH response in the case of short-medium length to the case of long helical peptides. We have chosen as test system a generic coiled coil formed by two chains of 29 amino acids, as well as a real peptide of the same length. We have observed that our system is more stable near physiological conditions; extreme pH values lead to a decrease in the coiled-coil stability that is due to the perturbation of the charge balance in the system and the loss of interactions between the two chains.

Beside the practical relevance of this work, as coiled coils are relevant systems in bio-chemistry/-physics, the main result of this his study is the validation of our proposed pH-dependent coarse-grained methodology [12]. Given the positive outcome of our tests, we can conclude that the pH-dependent methodology is a rather universal strategy to deal with different pH environments, at the coarse-grained level, and that the particularities of the model (ie. the specific form or parameters of the interaction potential) can be efficiently adapted to specific situations following the principle of "consistency across the scales".

# Tables

TABLE I. Transition temperatures, $T_m^*$, (in reduced units) for single-chain simulations of different lengths, $l$, using the original model (optimized for short chains).

| $l$ | $T_m^*$ |
|---|---|
| 13 | 2.3 |
| 15 | 2.3 |
| 17 | 2.2 |
| 19 | 2.2 |
| 21 | 2.2 |
| 23 | 2.1 |
| 25 | 2.3 |
| 27 | 2.5 |
| 29 | 2.5 |
| 31 | 2.2 |
| 33 | 2.2 |
| 35 | 2.3 |

TABLE II. Properties of a generic peptide in folded conditions at different pH values: transition temperature, $T_m^*$; proportion of coiled coils, $\%_{cc}$; relative ellipticity, $[\theta_{rel}]$, electrostatic energy in internal units, $E^{elec}$.

| pH | $T_m^*$ | $\%_{cc}$ | $[\theta_{rel}]$ | $E^{elec}$ |
|---|---|---|---|---|
| 1  | 2.8 | 0.78 | 0.84 | -0.83 |
| 7  | 3.1 | 1.00 | 0.86 | -1.51 |
| 13 | 3.0 | 0.32 | 0.82 | -0.02 |

# Figures

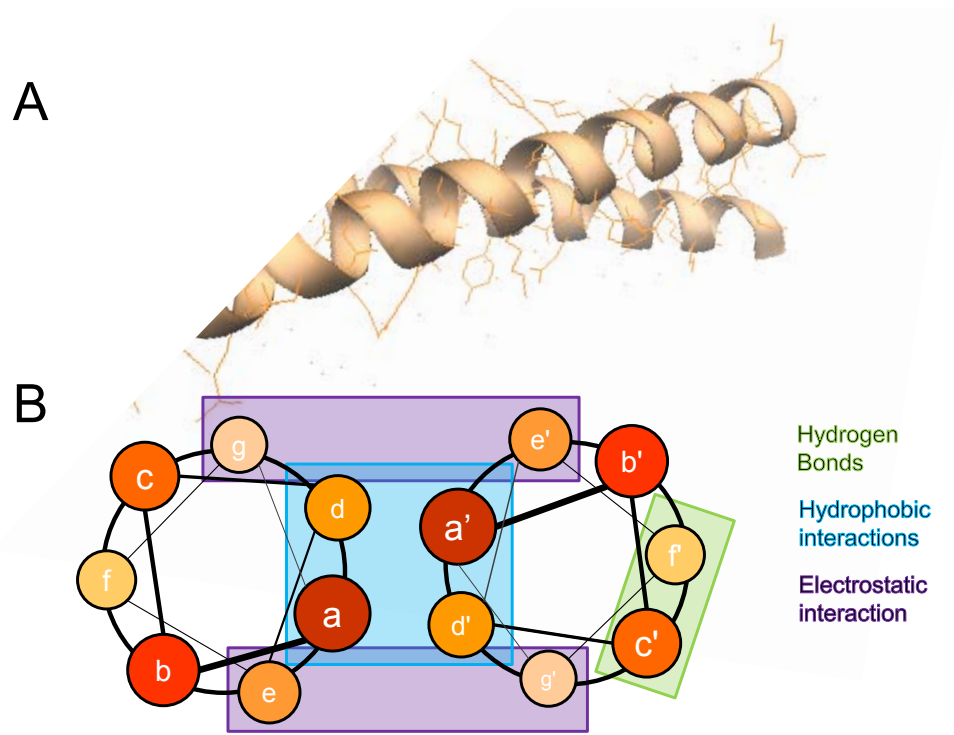

FIG. 1. Schematic representation of a coiled coil. A. Cartoon representation of a coiled-coil, where α-helices have been colored in purple. B. Upper view of a coiled coil, with labels in the coiled-coil heptad and colored boxed that represent the different interactions that are involved.

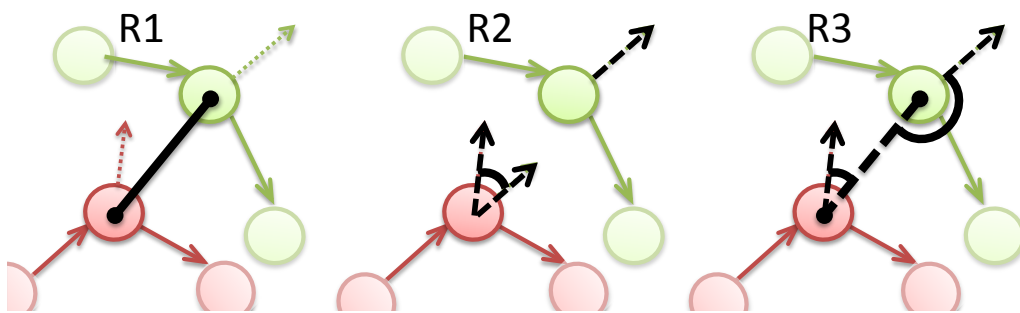

FIG. 2. Chain representation using the $C_\alpha$ resolution used in this work. Each bead represents one amino acid and the orange and green colors indicate two different chains. Besides, the hydrogen bond restrictions R1, R2 and R3 have been drawn. See text for details.

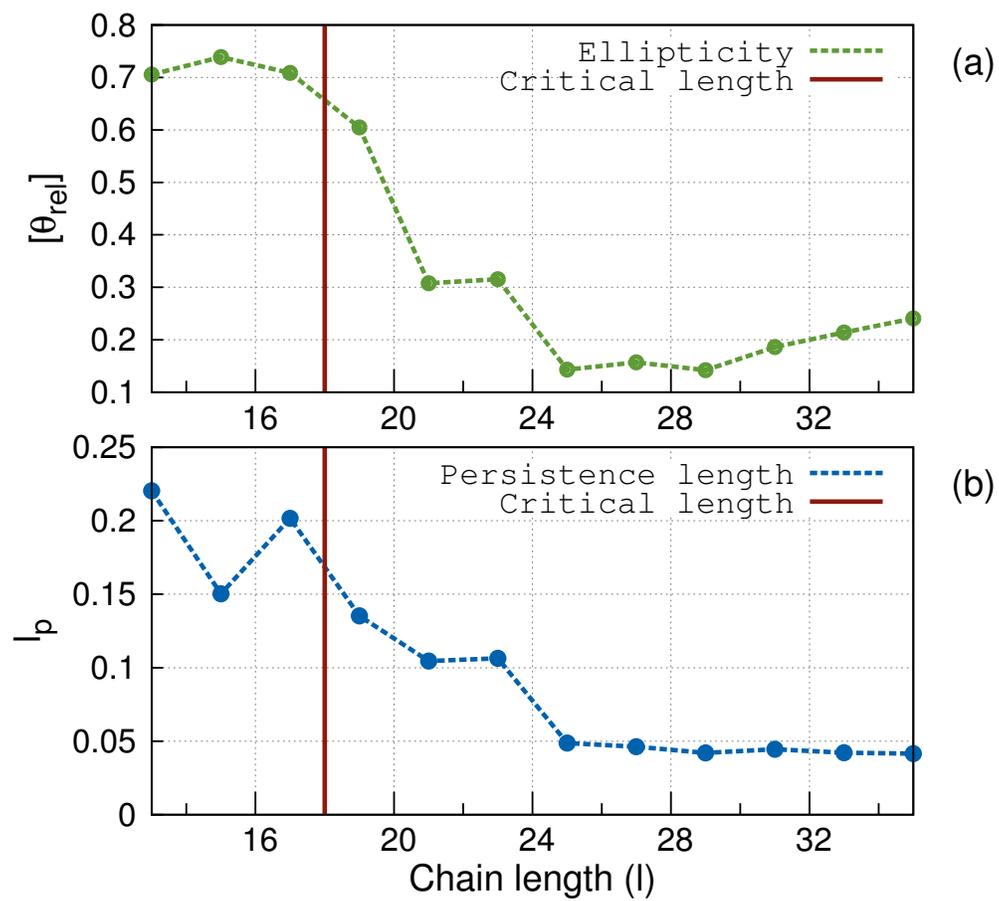

FIG. 3. Properties of chains of different chain lengths using the original interaction potential. (a) Relative ellipticity. (b) Persistence length. We have marked the critical length value with a red vertical line.

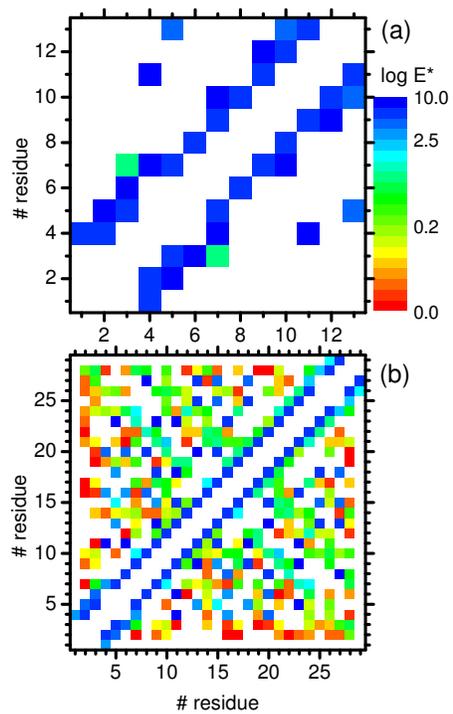

FIG. 4. Energetic maps using the original potential for two different chain lengths. (a) 13 residues (short chain). (b) 29 residues (long chain).

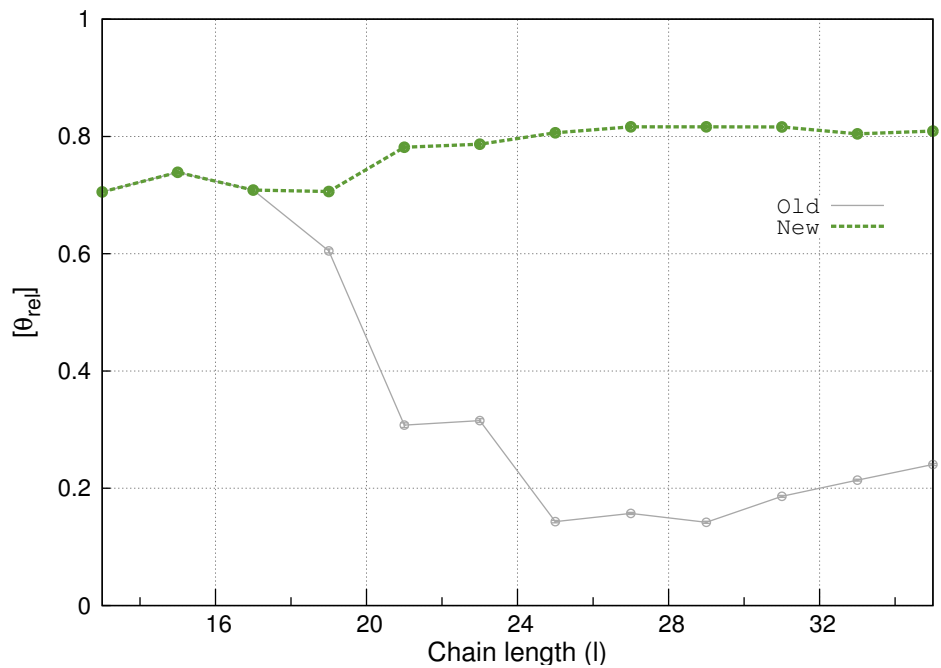

FIG. 5. (a) Relative ellipticity of chains of different chain lengths using the refined interaction potential. The old values have been plotted in grey.

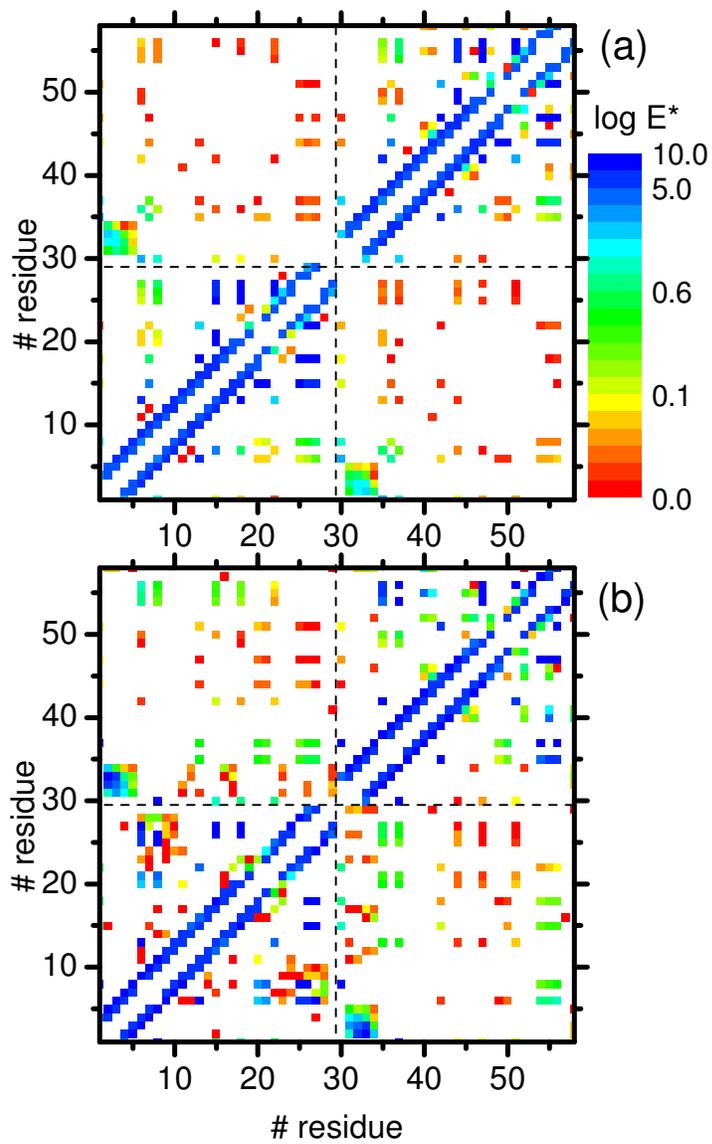

FIG. 6. Energetic maps of the B-ZIP domain at two pH values. (a) pH=3 (acid). (b) pH=7 (neutral).